\documentclass[12pt,preprint]{aastex}

\newcommand{\nltt}{NLTT~11748}

\newcommand{\lc}{light curve}
\newcommand{\lcs}{light curves}

\newcommand{\fov}[2]{\ensuremath{ \rm #1.\!^\prime #2 \times \rm #1.\!^\prime #2}}

\newcommand{\pxs}[2]{\ensuremath{#1.\!^{\prime\prime}#2\ \rm{pixel}^{-1}}}
\newcommand{\bin}[1]{\ensuremath{ \rm #1 \times \rm #1} \rm{pixel}}

\newcommand{\msun}{\ensuremath{M_\sun}}

\newcommand{\figr}[1]{Fig.~\ref{fig:#1}}
\newcommand{\secr}[1]{\mbox{\S\ \ref{sec:#1}}}
\newcommand{\eqr}[1]{Eq.~(\ref{eq:#1})}

\newcommand{\tabr}[1]{\mbox{Table~\ref{tab:#1}}}

\slugcomment{\it Submitted: 2010 September 18, Accepted: 2010 November 17}

\shorttitle{Relativistic beaming in a WD binary}

\shortauthors{Shporer et al.}

\begin{document}

\title{A ground-based measurement of the relativistic beaming effect in a detached double WD binary}

\author{Avi Shporer\altaffilmark{1, 2}, 
David L.~Kaplan\altaffilmark{3,4,5}, 
Justin D.~R.~Steinfadt\altaffilmark{2},
Lars Bildsten\altaffilmark{2,3}, 
Steve B.~Howell\altaffilmark{6}, 
Tsevi Mazeh\altaffilmark{7}} 

\altaffiltext{1}{Las Cumbres Observatory Global Telescope Network, 6740 Cortona Drive, Suite 102, Santa Barbara, CA 93117, USA; ashporer@lcogt.net}
\altaffiltext{2}{Department of Physics, Broida Hall, University of California, Santa Barbara, CA 93106, USA}
\altaffiltext{3}{Kavli Institute for Theoretical Physics, Kohn Hall, University of California, Santa Barbara,  CA 93106, USA}
\altaffiltext{4}{Hubble Fellow}
\altaffiltext{5}{Current address: Department of Physics, University of Wisconsin-Milwaukee, P.O. Box 413, Milwaukee, WI 53201 USA}
\altaffiltext{6}{National Optical Astronomy Observatory, 950 North Cherry Avenue, Tucson,  AZ 85719, USA}
\altaffiltext{7}{Wise Observatory, Tel Aviv University, Tel Aviv 69978, Israel}

\begin{abstract}

We report on the first ground-based measurement of the relativistic beaming effect (aka Doppler boosting). We observed the beaming effect in the detached, non-interacting eclipsing double white dwarf (WD) binary \nltt. Our observations were motivated by the system's high mass ratio and low luminosity ratio, leading to a large beaming-induced variability amplitude at the orbital period of 5.6 hr. We observed the system during 3 nights at the 2.0m Faulkes Telescope North with the SDSS-$g^\prime$ filter, and fitted the data simultaneously for the beaming, ellipsoidal and reflection effects. Our fitted relative beaming amplitude is $(3.0 \pm 0.4) \times 10^{-3}$, consistent with the expected amplitude from a blackbody spectrum given the photometric primary radial velocity amplitude and effective temperature. This result is a first step in testing the relation between the photometric beaming amplitude and the spectroscopic radial velocity amplitude in \nltt\ and similar systems. We did not identify any variability due to the ellipsoidal or reflection effects, consistent with their expected undetectable amplitude for this system. Low-mass, helium-core WDs are expected to reside in binary systems where in some of those systems the binary companion is a faint C/O WD and the two stars are detached and non-interacting, as in the case of \nltt. The beaming effect can be used to search for the faint binary companion in those systems using wide-band photometry. 

\end{abstract}

\keywords{stars: individual (NLTT~11748) --- white dwarfs}

\section{Introduction}
\label{sec:intro}

The movement of a light-emitting object relative to an observer causes relativistic beaming of the light (aka Doppler boosting) since the specific intensity is not a relativistic invariant.  When the object's radial velocity (RV) is periodically modulated, as is the case in binary systems, a periodic sinusoidal variation in the observed flux is induced.
This effect is one of three that can be seen in \lcs\ of binary systems, not necessarily eclipsing. The other two are the ellipsoidal effect, when a member of the binary system is tidally affected by the gravitational pull of its companion, and the reflection effect, when light originating from one component is reflected by the other.
The three effects were discussed in the context of planet-host stars by \cite{loeb03}, and main sequence stellar binaries by \cite{zucker07}. Both studies addressed binary systems with orbital periods much longer than a day, finding that the required photometric accuracy necessitates space-based observations.

Observational detections of the beaming effect are rare. \cite{maxted00} noted that  the beaming effect is likely to be seen in their ground-based data of a sdB star + WD binary system with a period of $P$=2.28 hr, although the effect's amplitude was not measured and the \lc\ is dominated by the ellipsoidal effect. \cite{vankerkwijk10} measured the beaming and ellipsoidal effects in a {\it Kepler} \lc\ of an A-type star + low-mass WD binary system with $P$=5.19 d \citep{rowe10}, and in fact used these measurements to resolve the puzzling nature of the system \citep[see also][]{Ehrenreich10}. \cite{mazeh10} identified the beaming and ellipsoidal effects in {\it CoRoT} data of a brown dwarf host star (CoRoT-3; $P$=4.26 d), and \cite{bloemen10} observed all three effects with {\it Kepler} in a 9.69 hr sdB + WD binary. 

We show here that for high mass ratio detached, non-interacting double white dwarf (WD) binaries the relative beaming amplitude is above the 10$^{-3}$ level, while the ellipsoidal and reflection amplitudes are smaller by an order of magnitude or more. Therefore, these systems represent an excellent opportunity to measure the beaming effect using ground-based telescopes. 

Following the discovery of the first eclipsing detached double WD binary \nltt\ by \cite{steinfadt10} we immediately noticed it is an excellent candidate for ground-based observations of the beaming effect. The observed RV amplitude is large \citep[$\approx$270 km s$^{-1}$;][]{steinfadt10, kawka10, kilic10b}, while the secondary is $\approx$30 times fainter than the primary \citep{steinfadt10}.
In addition, the orbital period is short enough to be covered within a single night. 

\nltt\ and similar systems where both the (photometric) beaming amplitude and the (spectroscopic) RV amplitude can be measured are a valuable tool for testing the relation between these two quantities, which is currently approximated using theoretical assumptions \citep[e.g.,][]{zucker07, vankerkwijk10}. 

We discuss the beaming effect in non-interacting double WD systems in \secr{wdbeaming}.
Our observations and data analysis are described in \secr{obs}. We present the results in \secr{res} where we also test them and compare our findings to theoretical expectations. In \secr{sum} we summarize our results and give our conclusions.

\section{Beaming in non-interacting double WDs}
\label{sec:wdbeaming}

The relatively large beaming amplitude in non-interacting double WD binaries originates from the unique WD mass-radius, or mass-luminosity relation. In those binary systems the photometric primary (hereafter, the primary) is a helium core WD, a few times smaller in mass but larger in radius and luminosity than the photometric secondary (hereafter, the secondary), a C/O core WD (see also Table 1 in \citealt{kaplan10}). As shown by \cite{zucker07} the observed beaming effect is the weighted difference between the individual beaming effects of the two binary components, shifted by a phase of 0.5. However, the larger mass and smaller luminosity of the C/O WD make its own beaming effect much smaller, and even negligible relative to that of the He WD. This can be seen by assuming both stars radiate as blackbodies and rewriting Equation~(6) of \cite{zucker07} as:
\begin{equation}
\label{eq:beamamp1}
A_{beaming} = \alpha^{'}_1\    \frac{K_1}{c}\    
 \frac{1-\frac{\alpha^{'}_2}{\alpha^{'}_1} \frac{m_1}{m_2} \frac{F_{\nu,2}}{F_{\nu,1}}}{1+F_{\nu,2}/F_{\nu,1}} \  , 
\end{equation}
where $K_1$ is the radial velocity amplitude of the primary, $c$ the speed of light, $m_1$ ($m_2$) is the mass of the primary (secondary) and $F_{\nu,1}$ ($F_{\nu,2}$) is the flux from the primary (secondary). $\alpha^{'}_{1,2}$ relates to the spectral index $\alpha_{1,2} \equiv \frac{d\log(F_{\nu,1,2})}{d\log(\nu)}$ as:
\begin{equation}
\alpha^{'}_{1,2} = 3 - \alpha_{1,2} = \frac{{x_{1,2}}\ e^{x_{1,2}}}{e^{x_{1,2}} - 1},
\end{equation}
where $x_{1,2} = h \nu / kT_{\rm eff,1,2}$ and $T_{\rm eff,1}$ ($T_{\rm eff,2}$) is the effective temperature of the primary (secondary).

For a high mass ratio non-interacting double WD binary the term $ \frac{1-\frac{\alpha^{'}_2}{\alpha^{'}_1} \frac{m_1}{m_2} \frac{F_{\nu,2}}{F_{\nu,1}}}{1+F_{\nu,2}/F_{\nu,1}}$ in \eqr{beamamp1} right hand side becomes of order unity. This follows from the mass ratio and the flux ratio of the two stars, and also from the fact that $\alpha^{'}_2 / \alpha^{'}_1$ is itself close to unity for the expected effective temperatures of the two stars. Therefore \eqr{beamamp1} reduces to:
\begin{equation}
\label{eq:beamamp2}
A(\nu)_{beaming} \approx\ \alpha^{'}_{1}\ \frac{K_1}{c}\ =\ 10^{-3}\ \alpha^{'}_{1}\ \frac{K_1}{300\ {\rm km\ s^{-1}}} \ ,
\end{equation}
or in physical parameters:
\begin{equation}
\label{eq:beamamp3}
A(\nu)_{beaming} \approx\ 2 \times 10^{-3}\ \alpha^{'}_{1}\ \frac{m_2}{\msun}\   \left(\frac{m_1+m_2}{\msun} \right)^{-2/3}\  \left(\frac{P}{\rm 1\ hour} \right)^{-1/3}\  \sin(i)  \  ,
\end{equation}
where $i$ the orbital inclination angle. The approximation neglects the contribution of the secondary and assumes a blackbody spectrum and a circular orbit. Since $\alpha^{'}_{1}$ has a value of a few in the optical for the typical temperatures of He WDs \citep[e.g.,][]{kilic10a}, the equation above shows that the beaming amplitude is expected to be at the few 10$^{-3}$ level, and therefore detectable from the ground. We note that this is not the accuracy required from a single measurement but the overall relative sinusoidal variability amplitude along the entire orbit.

\begin{figure}
\begin{center}
\includegraphics[scale=0.70]{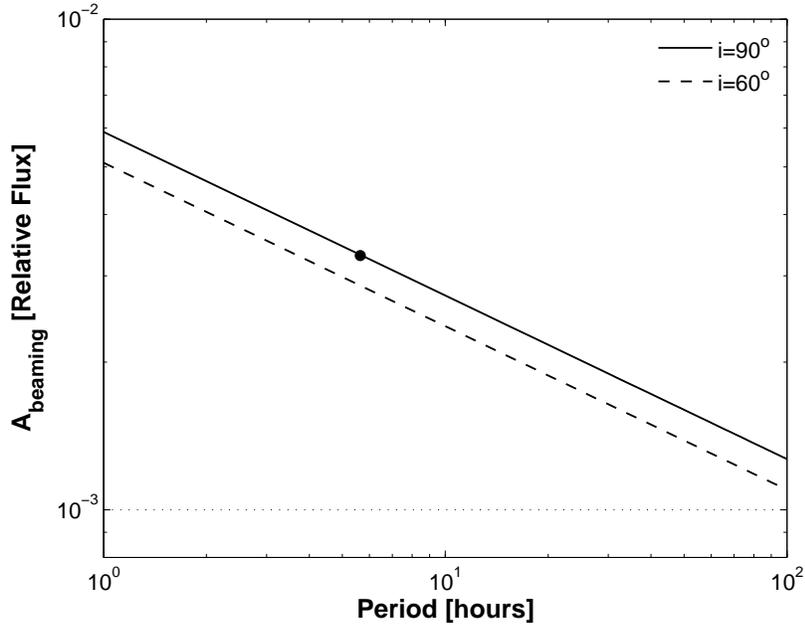}
\caption{\label{fig:beamamp2} Expected beaming amplitude according to \eqr{beamamp3}, for an \nltt-like system as a function of orbital period in the SDSS-$g^\prime$ band. The plot was produced using the masses of the two WDs in \nltt\ and the effective temperature of the primary. The solid line corresponds to a completely edge-on system, close to \nltt\ inclination of $89.9\pm0.1$ degrees, while the dashed line corresponds to a similar system with $i$=60 degrees. The black filled circle marks the expected \nltt\ position.
The plot clearly shows that similar systems to \nltt, with a longer period and a lower inclination are also expected to show a beaming amplitude above the 10$^{-3}$ level, marked with a horizontal dotted line.
}
\end{center}
\end{figure}

\figr{beamamp2} shows the expected SDSS-$g^\prime$ beaming amplitude versus orbital period, for \nltt-like systems (solid line). Also plotted (dashed line) is the expected amplitude for a system with an inclination angle of $i=60$ degrees. The figure shows that systems similar to \nltt, but with longer periods and/or lower inclinations are also expected to show a beaming amplitude above the 10$^{-3}$ level.

\section{Observations and Data Analysis}
\label{sec:obs}

We observed \nltt\ for 3.5--4.0 hr during each of the three nights of 2010 February 7, 9 \& 12 UT, obtaining data over all orbital phases. Most orbital phases were observed during two different nights. Our observations were done at the LCOGT\footnote{\url{http://lcogt.net}} 2.0m Faulkes Telescope North (FTN), located at the Haleakala Observatory, Hawaii. We used the Spectral Instruments 600 Series camera and the SDSS-$g^\prime$ filter. The camera has a 4K $\times$ 4K back-illuminated Fairchild Imaging CCD with a \fov{10}{5} field of view (FOV). We used the default \bin{2} binning mode, with an effective pixel scale of \pxs{0}{304}. The telescope pointing was adjusted to have the guiding camera FOV include a suitable guide star. Our observations were done while defocusing the telescope, spreading the stellar PSF over more pixels and increasing the overall open shutter time relative to the camera dead time, mostly CCD readout time. The exposure time was 120 s, resulting in a median cycle time of 143 s.

Data taken during each night was processed the same way. Preliminary reduction, including bias and flat-field corrections was done using standard IRAF routines. We applied aperture photometry using IRAF/phot by using a few aperture sizes and eventually choosing a radius of 10 pixels ($3. \!^{\prime\prime}04$), resulting in the smallest scatter in the final \lcs\ of the target and similar brightness stars, within 1 mag from that of the target. The target's \lc\ was calibrated by dividing it by a comparison \lc\ --- a weighted average \lc\ of four comparison stars showing no variability, whether inter- or intra-night. This calibration included propagation of the errors of both the target and comparison \lcs. 

Out of the total of 286 exposures seven were taken completely or partially during the 3 min eclipses \citep{steinfadt10}, and were ignored in our analysis. Two additional clear outliers in the target's \lc\ were visually verified to result from a cosmic ray hitting within the photometric aperture. Therefore, a total of 277 exposures were used in our analysis. 

The error estimates of the calibrated \lc\ were directly based on the results of IRAF/phot.
We rescaled these errors to better reflect  the true noise in the data, by fitting a 3rd degree polynomial to the flux ratio \lcs\ of each night, and inflating the error bars to make the reduced $\chi^2$ equal one. The derived scaling factors were 1.83, 1.37 \& 1.53 for the three nights, Feb.~7, 9 \& 12, respectively. We note that this polynomial fit was done only for the purpose of rescaling the error bars.

\begin{figure}
\begin{center}
\includegraphics[scale=0.70]{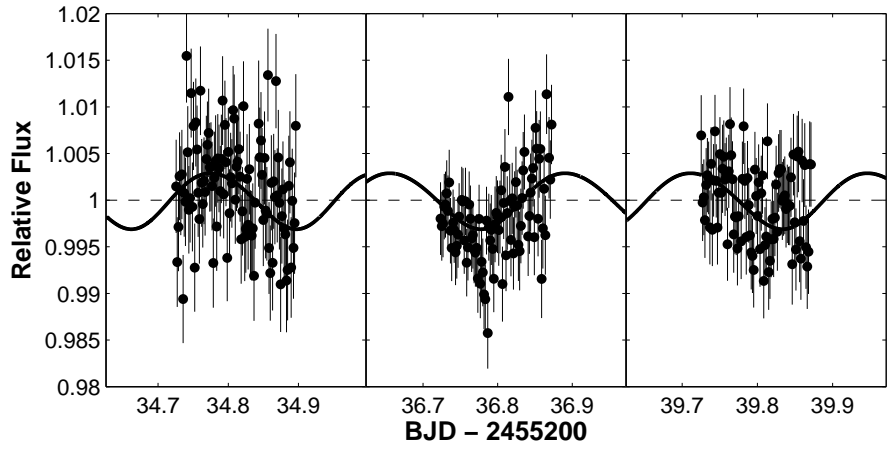}
\caption{\label{fig:lcnight} Nightly \lcs\ of \nltt, observed at FTN on the nights of (from left to right) 2010 February 7, 9 \& 12 UT. The overplotted solid line is the fitted model. Although the scatter in the data is relatively large, the small variability can be seen in the data.}
\end{center}
\end{figure}

The target's airmass varied from $\approx$1 to $\approx$2 during our observations, which is expected to affect the resulting \lc, depending on the difference in color between the target and comparison \lcs. We removed these trends by taking the data from all three nights together and fitting a 2nd degree polynomial to the target's relative flux versus airmass, and dividing by the fit. This step resulted in the final \lc, shown in \figr{lcnight}. 

In order to look for the beaming effect in the final \lc\ we simultaneously fitted for the three known effects (beaming, reflection and ellipsoidal). In our modeling we took the beaming effect to be a {\it sine} modulation at the orbital period. The reflection effect has the same period but is lagging in phase by $\pi/2$, hence it is a {\it cosine} modulation at the orbital period. The ellipsoidal effect has half the orbital period but the same phase as the reflection effect, so it is a cosine modulation at the first harmonic of the period. We used the ephemeris of \cite{steinfadt10} and fitted the phase folded \lc\ with a 5-parameter model \citep{sirko03, mazeh10}, including a sine and cosine terms for the orbital period and its first harmonic:
\begin{equation}
\label{eq:model}
f(t) = a_0 + a_{1c}\cos\left(\frac{2\pi}{P}\hat{t}\right) + a_{1s}\sin\left(\frac{2\pi}{P}\hat{t}\right) + a_{2c}\cos\left(\frac{2\pi}{P/2}\hat{t}\right) + a_{2s}\sin\left(\frac{2\pi}{P/2}\hat{t}\right)  , 
\end{equation}
where $\hat{t} \equiv t - T_{\rm 0}$, time subtracted the mid primary eclipse time. The ephemeris of  \cite{steinfadt10} is based on observations done during 1.5 months prior to the observations presented here. This resulted in a negligible uncertainty in the phase during our observations, smaller than 20 seconds, or 10$^{-3}$ of the period, justifying keeping the phase fixed in our fit.

\section{Results}
\label{sec:res}

\begin{figure}
\begin{center}
\includegraphics[scale=0.70]{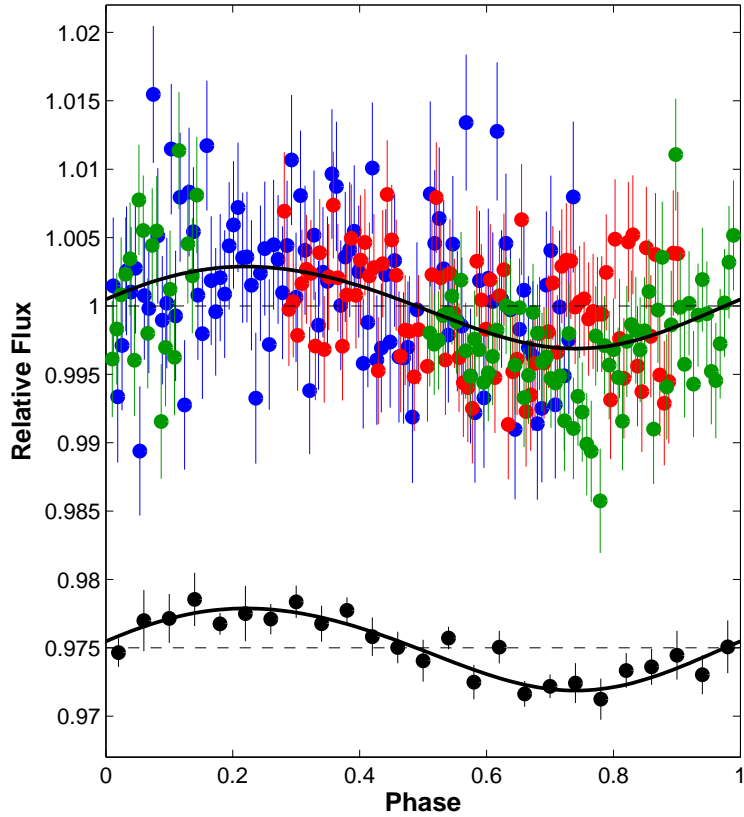}
\caption{\label{fig:lccomb} \nltt\ \lc\ phase folded with the known ephemeris. Phase zero is time of primary eclipse. Data from different nights are marked by different colors (top \lc). The bottom \lc\ was generated by binning the points in phase, using 0.04 phase bins, and was shifted downward for clarity. The binned \lc\ is shown for visualization only, and was not used in our analysis. Overplotted solid line (black) is the fitted 5-parameter model, constructed from the coefficients listed in \tabr{res}.}
\end{center}
\end{figure}

Results for the fitted coefficients are listed in the middle column of \tabr{res}. The fitted beaming relative amplitude is $a_{1s} = (3.0 \pm 0.4) \times 10^{-3}$, showing that the beaming signal is detected with a significance of $> 7\sigma$. The other three fitted coefficients were found to be consistent with zero. The $\chi^2$ statistics of the fitted model was 283.9, and the reduced $\chi^2$ was $\chi^2_{\rm red} \equiv \chi^2/{\rm dof} = 1.04$ for the 272 degrees of freedom, indicating a good estimate of the errors. \figr{lccomb} presents the phase folded \lc\ and the fitted model is overplotted.

\subsection{Further Tests}
\label{sec:test}

We have tested our results in order to verify they are not a statistical fluctuation, 
and that the beaming signal was not accidently injected into the data by our analysis.

Assuming the \lc\ is non-variable (equivalent to fitting a constant value equal to the weighted average) results in $\chi^2 = 351.5$ for 276 degrees of freedom, and $\chi^2_{\rm red} = 1.27$, clearly rejecting the no variability hypothesis. A bootstrap analysis, done by randomly permuting the \lc\ points in time showed that the probability of the measured beaming signal being a random one is smaller than 10$^{-6}$. Both these tests show, in addition to the amplitude's high S/N, that the detected variability has a high statistical significance.

It could be that the way we accounted for the airmass induced variation has injected the sine signal into the data. We tested this in two ways. First, we changed the polynomial degree used in fitting the relative flux versus airmass. Using a degree of 1 gave a 0.5$\sigma$ increase in the fitted beaming amplitude and increasing the polynomial degree up to 6 decreased the amplitude by $\leq 0.5\sigma$. In all these fits the other three fitted amplitudes were found to be consistent with zero. Second, we examined the \lcs\ of similar color stars in the field, with a $B-V$ term within 0.1 mag from that of \nltt. 
For those stars $\chi^2_{\rm red}$ was similar at the 0.01 level to that of a non-variable model, indicating they show no significant variability, and that the way we accounted for the airmass variation during our observations did not induce the detected variability.

Another possibility is that the detected signal originated from a variability in one of the four comparison stars. We checked this in two ways. First, we repeated our analysis four times while excluding one of the four comparison stars, using only the other three. This changed the fitted coefficients by less than 1$\sigma$ and resulted in similar $\chi^2$ values. Second, we applied the same analysis for each comparison star, using only the other three as comparisons. As in the case of the similar color stars described above, for all four comparison stars this resulted in a similar $\chi^2$ when fitting the 5-parameter model and a non-variable model, and the fitted beaming amplitude was consistent with zero. These two tests show that the variability identified in the target did not originate from the comparison stars.

We repeated our analysis three times using data from only two of the three nights. We were able to detect the beaming amplitude in each of the three analyzes, although with a lower significance, but still higher than 4$\sigma$. In each of those three analyzes the other three coefficients were found to be consistent with zero.

Finally, we fitted the data also with a simpler, 2-parameter model, including a zero point and only the sine component corresponding to the beaming amplitude ($a_{1s}$). The result was the same as in the original 5-parameter model, and is listed on the second column from the right in \tabr{res}.

\subsection{Comparison with expectations}
\label{sec:comp}

For estimating the expected amplitude of the beaming effect in our data we first calculated it as a function of wavelength assuming a blackbody spectrum, using \eqr{beamamp2}. This function is plotted as a solid black line in \figr{beamamp}. We then weighted it by the SDSS-$g^\prime$ filter transmission curve and the CCD QE. The final estimates for each band are marked as filled circles in \figr{beamamp} where the error bars due to the uncertainties in $T_{\rm eff,1}$ and $K_1$ are comparable to the size of the markers. \figr{beamamp} also justifies our choice to observe in SDSS-$g^\prime$, as the beaming amplitude increases towards the blue, but observations in SDSS-$u'$ are difficult at FTN. The expected SDSS-$g^\prime$ amplitude is ($3.3 \pm 0.1) \times 10^{-3}$, which is within 1$\sigma$ from the measured amplitude of ($3.0 \pm 0.4) \times 10^{-3}$, marked by a diamond in \figr{beamamp}. Therefore, the measured amplitude for the beaming effect is consistent with the expected value assuming a blackbody spectrum.

Another way to estimate the expected beaming amplitude, in a more direct manner and without the need to assume a blackbody spectrum is to integrate over a measured spectrum of the system using Eq.~2 of \cite{vankerkwijk10}. For that end we used the NTT/EFOSC2 spectrum obtained by \cite{kawka09}, which covers the SDSS-$g^\prime$ band. This resulted in an expected amplitude of $3.7 \times 10^{-3}$, less than 2$\sigma$ away from the measured amplitude. However, we caution that the uncertainty on this result is difficult to estimate. This low resolution spectrum was derived from two 20 min exposures, and was affected by the Doppler shift variation during the exposure, and by the RV difference between the two exposures, which could have been at the $\sim$50 km s$^{-1}$ level, depending on the exact orbital phases they were taken at.

We estimate the amplitude of the ellipsoidal effect following \cite{morris93} to be about $\frac{m_2}{m_1}\left(\frac{r_1}{a}\right)^3$, where $a$ is the orbital semi-major axis, $r_1$ the radius of the primary, and we have ignored an order to unity coefficient \citep[see also][]{loeb03, zucker07, mazeh10}. This expression is $< 10^{-4}$ for \nltt.
Although it is only a rough estimate and a more accurate one can be derived for this system, we use it only to show that the ellipsoidal effect is too small to be detected in our data.
For the reflection effect we use $ \left( \frac{r_2}{a} \right)^2$ as an upper limit, which is $4 \times 10^{-5}$ for \nltt. Therefore, both the ellipsoidal and reflection effects are not expected to be detected from the ground, in agreement with our results.

\section{Summary and Conclusions}
\label{sec:sum}

We have presented a ground-based measurement, carried out with a 2.0m telescope, of the relativistic beaming effect in the high mass ratio WD binary \nltt\ \citep{steinfadt10}. We showed that the measured amplitude, of $(3.0 \pm 0.4) \times 10^{-3}$, is within 1$\sigma$ from an approximated theoretical estimate, and within 2$\sigma$ from a more direct estimate using a low resolution spectrum of the system. The ellipsoidal and reflection effects were not detected in our data, in agreement with theoretical expectations. To our knowledge this is the first ground-based measurement of the relativistic beaming effect. 

Our measurement is a first step in testing the relation between the beaming and RV amplitudes in systems similar to \nltt, and it suggests it can be approximated by assuming a blackbody spectrum. Although we were not able to obtain a high precision measurement of the beaming amplitude, it is identified at a high significance, of more than 7$\sigma$. 

\begin{figure}
\begin{center}
\includegraphics[scale=0.70]{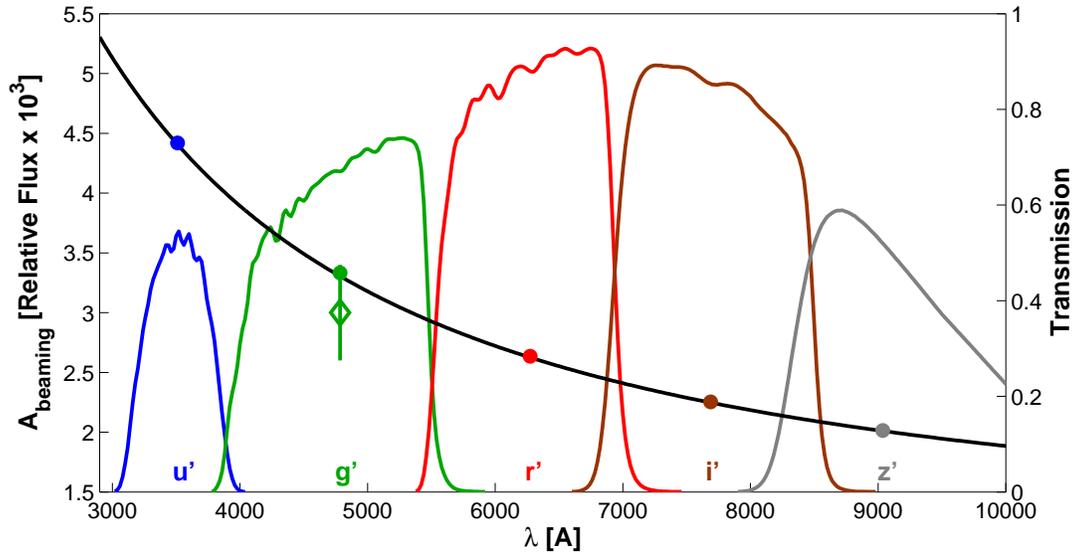}
\caption{\label{fig:beamamp} Amplitude of the expected relativistic beaming effect for \nltt\ as a function of wavelength (black solid line), assuming a blackbody spectrum. The filled circles are the calculated amplitudes in each filter, accounting for the CCD QE and filter transmission curves. The plotted colored curves account for both the CCD QE and filter transmission. The SDSS-$g^\prime$ beaming amplitude measured here is marked by a diamond, along with error bars (green).
}
\end{center}
\end{figure}

Although short period detached He + C/O WD binaries are rare, their discovery rate is increasing. Five such systems were recently discovered \citep{mullally09, kilic10a, marsh10, kulkarni10} by RV monitoring, and are expected to show beaming amplitudes of $\gtrsim 2 \times 10^{-3}$ in the optical. All five have orbital periods of only a few hours, shorter than that of \nltt, so they can be used to further test the relation between the RV amplitude and the beaming amplitude, and also the wavelength dependency of the latter, shown in \figr{beamamp} for \nltt. 

The evolution of a low-mass He WD requires it to have a binary companion \citep[e.g.,][]{marsh95}. Therefore, once a He WD is identified the beaming effect can be used to look for a binary companion photometrically, in a very similar way to spectroscopic monitoring. Since in this case the \lc\ becomes analogous to the RV curve, high precision photometry can in principle be used to model the orbit and measure the mass ratio of the two components. 

\acknowledgments
 
The authors warmly thank Marten van Kerkwijk and Shai Zucker for helpful discussions.
This paper uses observations obtained with facilities of the Las Cumbres Observatory Global Telescope.
DLK was supported by NASA through Hubble Fellowship Grant \#01207.01-A awarded by the STScI which is operated by AURA, Inc., for NASA, under contract NAS 5-26555.
This work was supported by the National Science Foundation under grants
PHY 05-51164 and AST 07-07633.

{\it Facilities:} \facility{FTN (Spectral)}


\clearpage

\begin{deluxetable}{ccccc}
\tablecaption{\label{tab:res} Fitted coefficients} 
\tablewidth{0pt}
\tablehead{& & \colhead{5-parameter model} &\colhead{2-parameter model} &\\
\colhead{Coefficient} & \colhead{Effect} & \colhead{Fitted value} &  \colhead{Fitted value} &\colhead{Expected value} \\
& & \colhead{x 10$^{3}$} & \colhead{x 10$^{3}$} & \colhead{x 10$^{3}$}  }
\startdata
$a_{1s}$ &  Beaming     &  $3.0 \pm 0.4$ & $3.0 \pm 0.4$ & $3.3  \pm 0.1$       \\ 
$a_{1c}$ &  Reflection   & $0.3 \pm 0.4$ & --- & $<0.1$  \\
$a_{2s}$ &  ---  & $0.1 \pm 0.4$ & --- & \multicolumn{1}{c}{---} \\
$a_{2c}$ &  Ellipsoidal  & $0.1 \pm 0.4$ & --- & $<0.1$  \\
\enddata
\end{deluxetable}

\clearpage

\clearpage

\clearpage

\end{document}